\documentstyle[11pt,newpasp,twoside,epsf]{article}
\def\edcomment#1{\iffalse\marginpar{\raggedright\sl#1\/}\else\relax\fi}
\reversemarginpar

\begin{document}
\title{Jodrell Bank Timing Astrometry}
 \author{G. Hobbs}
\affil{Jodrell Bank Observatory, Macclesfield, Cheshire, SK11 9DL, UK}
\affil{ATNF, CSIRO, PO Box 76, Epping
 NSW 1710, Australia}
\author{A.G. Lyne, M. Kramer}
\affil{Jodrell Bank Observatory, Macclesfield, Cheshire, SK11 9DL, UK}

\begin{abstract}
More than 500 pulsars are regularly observed with the 76-m Lovell
radio telescope at Jodrell Bank Observatory.  Precise positional and
rotational parameters have been obtained from observations spanning
between 6 and 34 years for over 300 of these pulsars.  The
parameters were determined by fitting a timing
model to whitened pulse arrival times. In this paper, 
the technique used to whiten the timing residuals is summarised and
the astrometric measurements are compared
to proper motions determined earlier with interferometers. This
novel approach has led to the first proper motion measurements for
111 pulsars and improved results for 15 pulsars.
\end{abstract}

\section{Introduction}

For pulsars with significant timing noise, proper motion measurements 
have traditionally been obtained using 
interferometers such as the VLA, VLBI or MERLIN
(for example, Brisken 2001; Lyne, Anderson \& Salter 1982 and Harrison, Lyne \&
Anderson 1993).  We show here that timing observations can provide a
simple method for obtaining the proper motions of a large number of
pulsars.  These results agree well with published values if timing noise is
correctly removed from the pulsars' timing residuals. The astrophysics obtainable 
from such a large sample of proper motion measurements is
manyfold. For example, the birth velocities may arise due to
asymmetric supernovae. The pulsar velocities are, in many cases,
greater than the escape velocities of binary systems, globular 
clusters and the Galaxy. The isotropic distribution of pulsars in 
the Milky Way may be distributed in a fashion similar to the sources 
producing $\gamma$-ray bursts and tracing a pulsar's position back in 
time can lead to new supernova remnant associations.

More than 500 pulsars are regularly observed using the 76-m Lovell
radio telescope at Jodrell Bank observatory.  For each observation,
folded pulse profiles, two orthogonal polarizations and a time stamp
are recorded.  Observations have been made predominately at
frequencies close to 408, 610, 910, 1410 or 1630\,MHz.  During 
standard observing procedures, the signals of each polarization are
mixed to an intermediate frequency, fed through a multi--channel
filterbank and digitized.  The data are dedispersed in hardware, 
folded on--line according to the pulsar's dispersion measure and
topocentric period and the polarizations added. 
Pulse times--of--arrivals (TOAs) are subsequently
determined by convolving the averaged profile with a template of high
S/N corresponding to the observing frequency.  The TOAs are corrected
to the solar system barycentre using the Jet Propulsion Laboratory
DE200 solar system ephemeris (Standish 1982). 

\section{FITWAVES}

The rotational and astrometric parameters for each pulsar are obtained
by model--fitting the TOAs using TEMPO\footnote{see http://pulsar.princeton.edu/tempo}.
A proper motion leads to a sinusoid, in the timing residuals, with a 
period of one year and an amplitude that 
increases linearly with time.  However, the timing residuals for many pulsars contain
timing noise, a continuous, noise like fluctuation in rotation rate, 
which must be removed before fitting the
timing model.  This removal, or whitening of the data, has 
traditionally been carried out by removing polynomials from the timing 
residuals.  However, with this method, the proper motion signature can be affected 
due to the lack of control over the frequency range of the timing 
noise spectrum being removed from the timing residuals.

\begin{figure}[!tb] \begin{center}
\plottwo{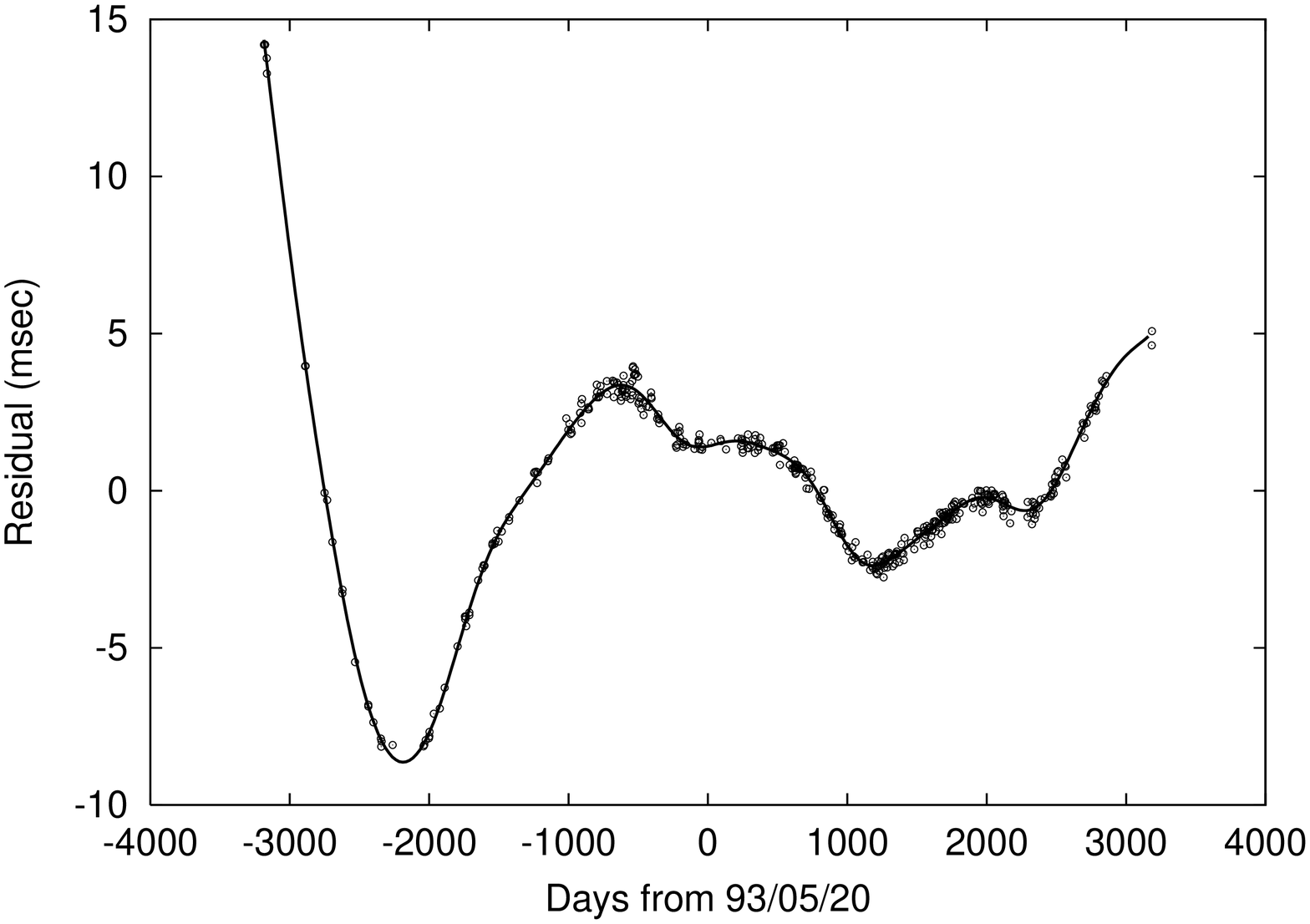}{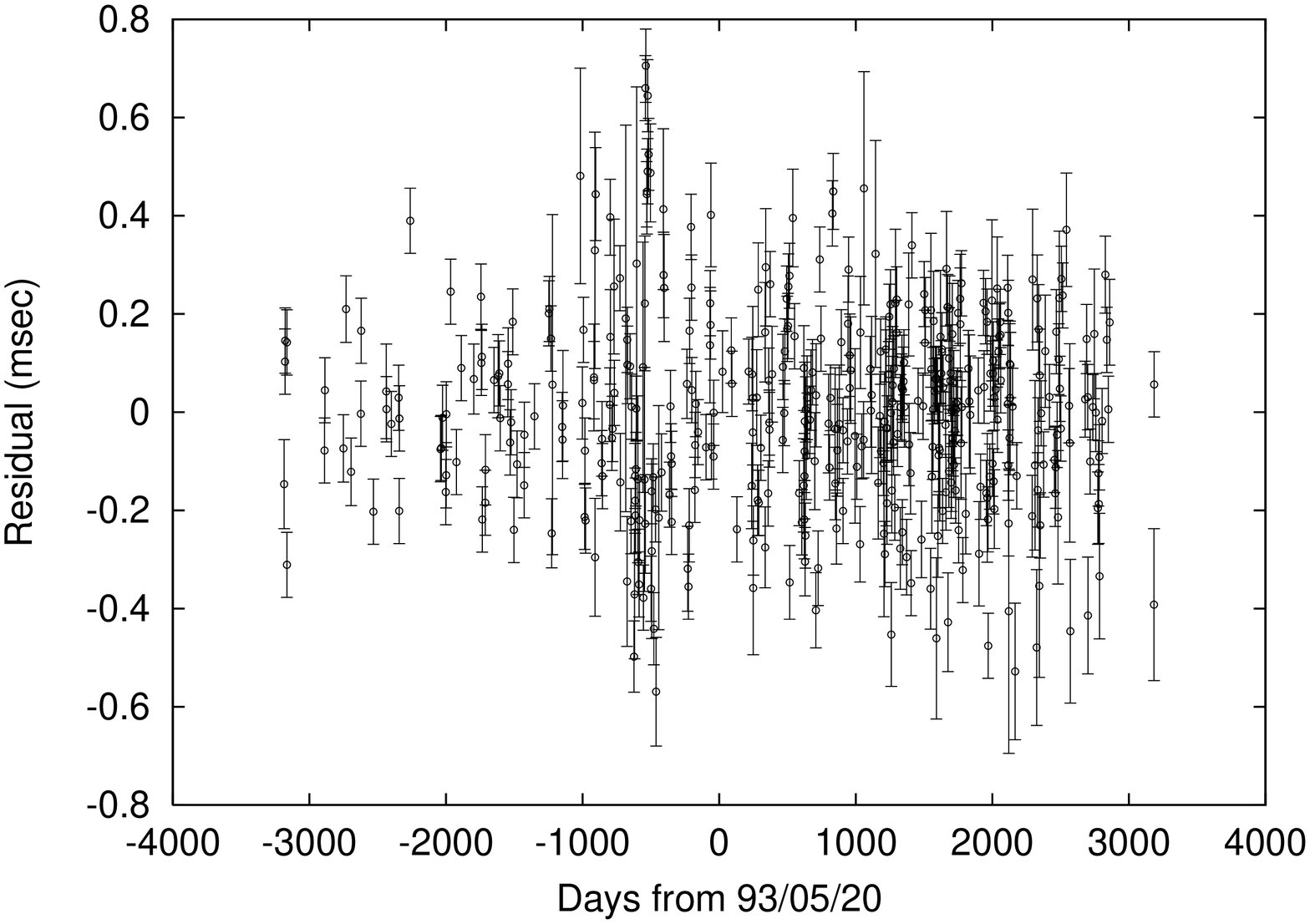}
\caption{The timing residuals for PSR~B0450$-$18, (a) pre- and (b)
post- whitening. }
\end{center} \end{figure}

A method, known as \textsc{fitwaves}, to whiten timing residuals
has been developed that fits harmonically related sinusoids to the
timing residuals. The number of harmonics used is determined by the
period of the shortest structure required to be removed from the
timing residuals. Therefore, \textsc{fitwaves} can remove long--period
timing noise without affecting the proper motion signature.
Figure 1 shows the technique being applied to the timing residuals 
of PSR~B0450$-$18.  After fitting to the pulse TOAs for rotational 
period and its first two derivatives, the residuals have the form 
shown in Figure 1a.  The curve fitted through the 
points has been obtained by fitting 
thirteen harmonically related sines and cosines to these residuals.
Whitened timing residuals are
formed by removing this curve from the original residuals (Figure
1b).  With these whitened residuals, it is possible to determine
the pulsar's proper motion in right ascension, $\mu_\alpha =
11(2)$\,mas\,yr$^{-1}$, and declination, $\mu_\delta =
6(4)$\,mas\,yr$^{-1}$ which can be compared to interferometric
measurements of $\mu_\alpha = 12(8)$ and $\mu_\delta
= 18(15)$\,mas\,yr$^{-1}$ published in Fomalont et al (1997).

\subsection{Comparison of proper motions}

 Pulsar astrometric parameters were determined for 321 pulsars 
that have been observed for more than six years and 
have no history of glitching.  The pulsars' positions, proper motions,
dispersion measures and derivatives were determined by fitting a
timing model to the whitened TOAs 
using \textsc{tempo}.  The data were whitened using the \textsc{fitwaves} method 
in such a way that timing noise features with periods greater than 
1.5 years were removed leaving any remaining structure unchanged.  

The proper motions for a large sample of pulsars have been tabulated 
in Hobbs et al. (2002) and will be described in detail in a subsequent paper
(Hobbs et al. in preparation).  This work has improved the precision 
to which 15 proper motions have been measured and has obtained the first 
proper motion results for 111 pulsars providing a sample of over 200 pulsars
with measured proper motions.  Here we show that the results 
being obtained for pulsars, whose timing residuals contain timing noise,  
agree with interferometric results.  

The most precise pulsar proper motions have been obtained by
Brisken (2001) using the VLBA.  Figure 2 compares seven pulsars that
were included in both analyses.  These results are clearly
consistent.  

\begin{figure} \begin{center}
\plottwo{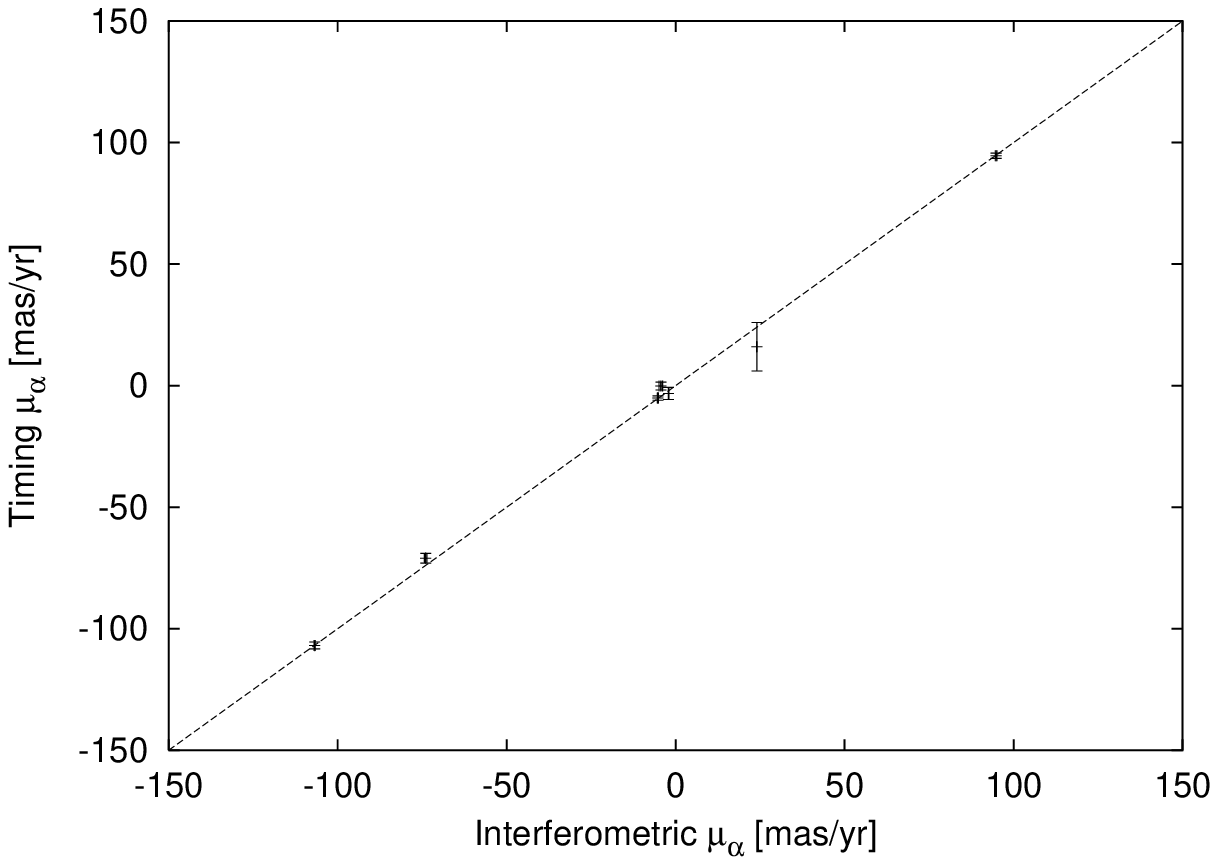}{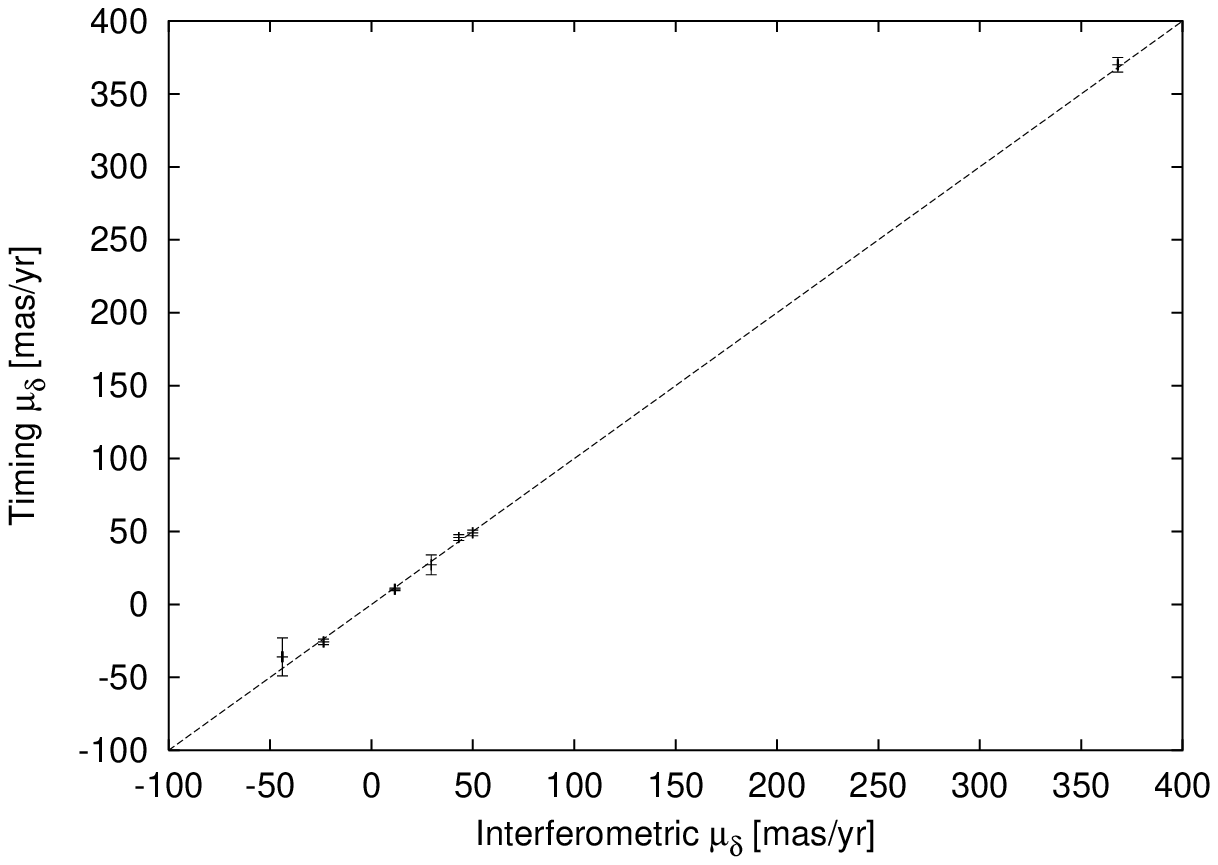}
\caption{The timing results obtained using the method described here
compared to the interferometric results of Brisken (2001).}
\end{center} \end{figure}

Earlier interferometric proper motions have been published using the VLA
(Brisken 2001; Fomalont et al. 1997) and the MERLIN arrays (Lyne,
Anderson \& Salter 1982; Harrison, Lyne \& Anderson 1993).  The most
precise interferometric results from these publications are compared,
in Figure 3, to the proper motions acquired using the timing method.
Although an inconsistency exists for the $\mu_\delta =
-98(6)$\,mas\,yr$^{-1}$ measurement of PSR~B0906$-$17 compared 
to $-40(11)$\,mas\,yr$^{-1}$ in
Harrison, Lyne \& Anderson (1993), the majority of the proper motions are
consistent with earlier results.

\begin{figure} \begin{center}
\plottwo{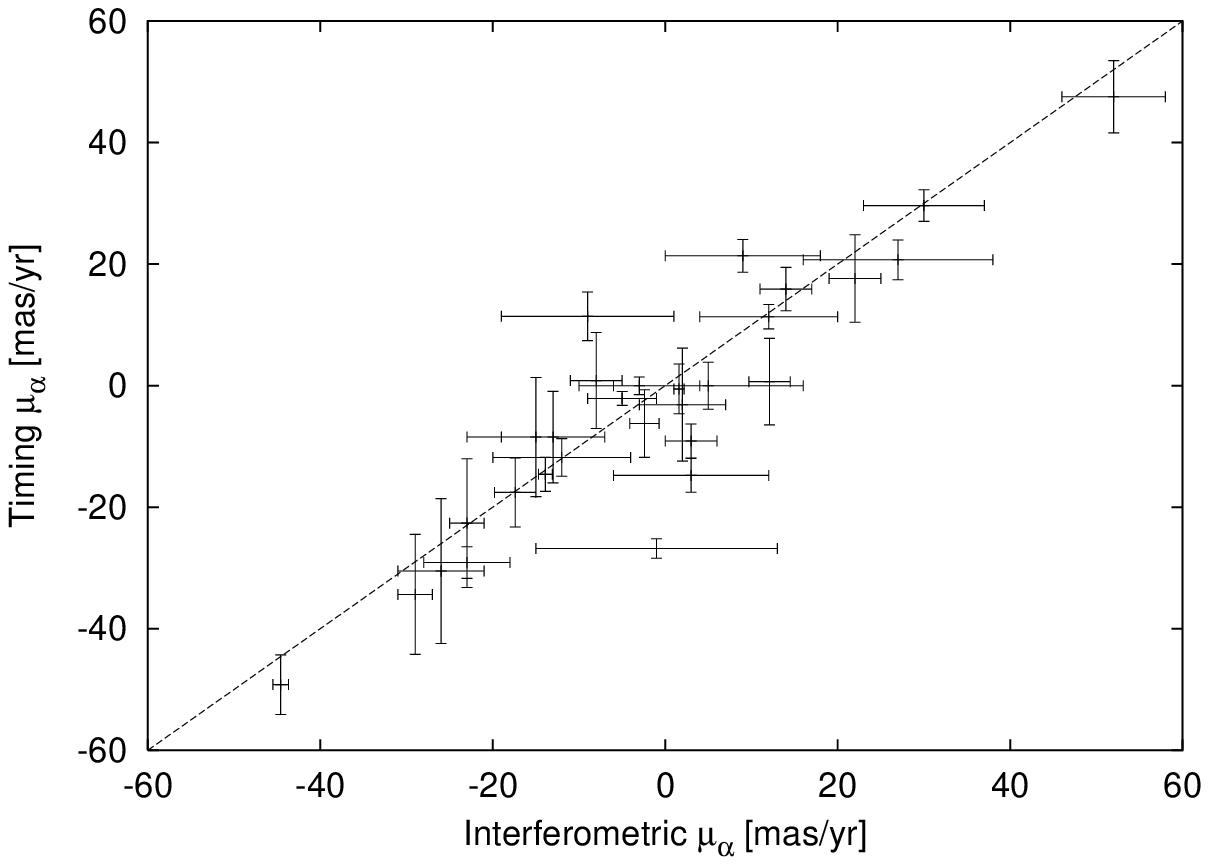}{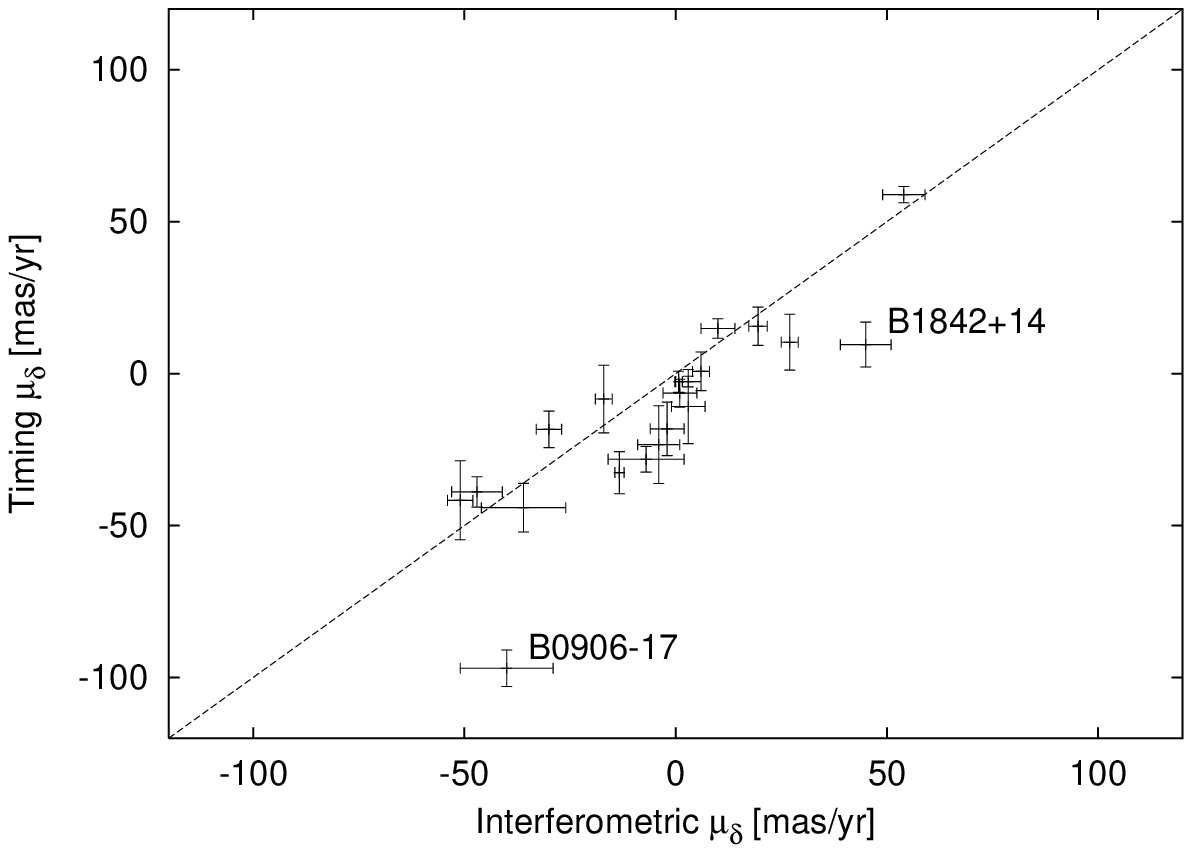}
\caption{The proper motions in right ascension ($\mu_\alpha$) and
declination ($\mu_\delta$) obtained using the method described here to
the MERLIN or VLA interferometric results.}
\end{center} \end{figure}

\section{Conclusion}

Pulsars' rotational and astrometric parameters can be obtained from
whitened timing residuals.   The timing noise present in the timing
residuals can be removed by fitting harmonically related sinusoids to
the data.  The astrometric parameters obtained from this timing
method agree with those measured using interferometers, even if timing
noise is present in the original non-whitened timing residuals.
Applying this technique to all the pulsar residuals  stored in the
Jodrell Bank data archive has led to more than 100 new proper motions
(and hence pulsar velocities) being measured.

\end{document}